\documentclass[12pt]{article}
\usepackage{amssymb,epsf,color}
\def\lsim{\mathrel{\rlap {\raise.5ex\hbox{$ < $}}
{\lower.5ex\hbox{$\sim$}}}}
\def\gsim{\mathrel{\rlap {\raise.5ex\hbox{$ > $}}
{\lower.5ex\hbox{$\sim$}}}} 
\def\sqr#1#2{{\vcenter{\vbox{\hrule height.#2pt

        \hbox{\vrule width.#2pt height#1pt \kern#1pt

           \vrule width.#2pt}

        \hrule height.#2pt}}}}

\def\lsim{{\displaystyle
{{\raise-8pt\hbox{$ <$}}
\atop{\raise5pt\hbox{$\sim$}}}}}
\def\gsim{{\displaystyle
{{\raise-8pt\hbox{$ >$}}
\atop{\raise5pt\hbox{$\sim$}}}}}
\def\slsim{{\displaystyle
{{\raise-8pt\hbox{$\scriptstyle <$}}
\atop{\raise5pt\hbox{$\scriptstyle \sim$}}}}}
\def\sgsim{{\displaystyle
{{\raise-8pt\hbox{$\scriptstyle  >$}}

\atop{\raise5pt\hbox{$\scriptstyle \sim$}}}}}
\newskip\humongous \humongous=0pt plus 1000pt minus 1000pt

\newcommand{\sumpf}[0]{\sum_{(H^{\rm f},G^{\rm f})}\! \! \! \!
{\raise
4pt
\hbox{$'$}}\,}

\newcommand{\sump}[0]{\sum_{(H,G)}\! \! {\raise 4pt \hbox{$'$}}\,}

\def\bs{\begin{subequations}}
\def\es{\end{subequations}}
\catcode`\@=11
\newcount\hour
\newcount\minute
\newtoks\amorpm

\hour=\time\divide\hour by60
\minute=\time{\multiply\hour by60 \global\advance\minute by-\hour}
\edef\standardtime{{\ifnum\hour<12 \global\amorpm={am}%
        \else\global\amorpm={pm}\advance\hour by-12 \fi

        \ifnum\hour=0 \hour=12 \fi
        \number\hour:\ifnum\minute<10 0\fi\number\minute\the\amorpm}}
\edef\militarytime{\number\hour:\ifnum\minute<10 0\fi\number\minute}
\def\draftlabel#1{{\@bsphack\if@filesw {\let\thepage\relax
   \xdef\@gtempa{\write\@auxout{\string
      \newlabel{#1}{{\@currentlabel}{\thepage}}}}}\@gtempa
   \if@nobreak \ifvmode\nobreak\fi\fi\fi\@esphack}
        \gdef\@eqnlabel{#1}}
\def\@eqnlabel{}
\def\@vacuum{}
\def\draftmarginnote#1{\marginpar{\raggedright\scriptsize\tt#1}}
\def\draft{\oddsidemargin -.2truein
        \def\@oddfoot{\sl preliminary draft \hfil
        \rm\thepage\hfil\sl\today\quad\militarytime}
        \let\@evenfoot\@oddfoot \overfullrule 3pt
        \let\label=\draftlabel
        \let\marginnote=\draftmarginnote
   \def\@eqnnum{(\theequation)\rlap{\kern\marginparsep\tt\@eqnlabel}%
\global\let\@eqnlabel\@vacuum}  }
\def\subequations{\refstepcounter{equation}%
  \edef\@savedequation{\the\c@equation}%
  \@stequation=\expandafter{\theequation}%   %only want \theequation
  \edef\@savedtheequation{\the\@stequation}% % expanded once
  \edef\oldtheequation{\theequation}%
  \setcounter{equation}{0}%
  \def\theequation{\oldtheequation\alph{equation}}}

\def\endsubequations{\setcounter{equation}{\@savedequation}%
  \@stequation=\expandafter{\@savedtheequation}%
  \edef\theequation{\the\@stequation}\global\@ignoretrue
  \vspace*{-12pt} \\}

\def\bs{\begin{subequations}}
\def\es{\end{subequations}}
\relax
\def\thefootnote{\fnsymbol{footnote}}
\def\be{\begin{equation}}
\def\ee{\end{equation}}
\def\ba{\begin{eqnarray}}
\def\ea{\end{eqnarray}}

\def\ee{\end{equation}}
\def\bea{\begin{eqnarray}}
\def\eea{\end{eqnarray}}

\newcommand{\uarrw}[0]{\mathrel{
{\raise.5ex\vbox{\hrule width 1cm}\hskip-6pt\rightarrow}}}
\def\thebibliography#1{%
\vskip 0.5cm \centerline{\bf References}
\list{%
[\arabic{enumi}]}{\settowidth\labelwidth{[#1]}
\leftmargin\labelwidth
\advance\leftmargin\labelsep
\usecounter{enumi}}
\def\newblock{\hskip .11em plus .33em minus .07em}
\sloppy\clubpenalty4000\widowpenalty4000
\sfcode`\.=1000\relax}

\renewcommand{\theequation}{\arabic{section}.\arabic{equation}}

\renewcommand{\section}{\setcounter{equation}{0}\@startsection%
{section}{1}{0mm}{-\baselineskip}{0.5\baselineskip}%
{\normalfont\normalsize\bfseries}}

\renewcommand{\subsection}{\@startsection%
{subsection}{2}{0mm}{-\baselineskip}{0.5\baselineskip}%
{\normalfont\normalsize\slshape}}

\renewcommand{\subsubsection}{\@startsection%
{subsubsection}{2}{0mm}{-\baselineskip}{0.5\baselineskip}%
{\normalfont\normalsize\slshape}}

\begin{document}
%
%\special{!userdict begin /bop-hook{gsave 200 30 translate
%65 rotate /Times-Roman findfont 216 scalefont setfont
%0 0 moveto 0.85 setgray (\jobname) show grestore}def end}
% 
%\renewcommand{\theequation}{\arabic{section}.\arabic{equation}}
\renewcommand{\theequation}{\arabic{equation}}
\begin{titlepage}
\begin{flushright}

\end{flushright}
\begin{centering}
\vspace{1.0in}

{ \large \bf Beyond Quantum Mechanics and General Relativity} $^1$\\

\vspace{2.5 cm}

{\bf Andrea Gregori} $^2$\\
\medskip
\vspace{2.8cm}
{\bf Abstract} \\
\end{centering} 
\vspace{.2in}
In this note I present the main ideas of my
proposal about the theoretical framework that could underlie, and
therefore ``unify'', Quantum
Mechanics and Relativity, and I briefly summarize 
the implications and predictions.

\vspace{9cm}

\hrule width 6.7cm
\noindent
$^1$ To appear in the ``Invertis Journal of Science and Technology'', 
New Dehli, 2010.
\newline
$^2$
e-mail: agregori@libero.it

\end{titlepage}
\newpage
\setcounter{footnote}{0}
\renewcommand{\thefootnote}{\arabic{footnote}}

The entire last century of physics is marked
by two main theories: Quantum Mechanics, and Relativity.
Implementation of Relativity in Quantum Mechanics has produced the branch 
of Field Theory.
It is widely accepted that, as it happens for electromagnetism,
and the weak and strong forces, at a certain scale also gravity should
be quantized, and that perhaps all
fundamental forces should be ``unified'' in a quantum theory of gravity.

Various progresses have
been made in the study of string theory, considered the most promising
candidate to be the theory satisfying this kind of theoretical expectations. 
However, until now ``the'' solution still seems to
remain elusive, because string theory is built as a
``free'' theory, seemingly allowed
to take any form one wants to force on it (apart from the right one!). 
Indeed, as it is by now so common to think in terms of ``quantization'' of
a theory, it is however not as much clear, 
even for the cases in which the quantum nature is out of discussion,
\emph{why} physical phenomena do show a quantum behaviour, and therefore also
why should we expect to find quantum aspects also in gravity, apart from 
its loose analogy with the other forces \footnote{There have been
attempts to derive Quantum Mechanics within the framework of 
Classical Mechanics, but these, apart from providing an interesting mapping 
of the problem into the formalism of statistical theory,
don't solve the main question.}.
Similarly, also the fact that the speed of light
has a universal bound, $c$, a clear experimental observation, is assumed, 
rather than derived, in the theoretical formulation that implements
this phenomenon; namely, the Einstein's Theory of Relativity.

All these aspects, namely, the fact that 
we don't know why nature is quantized, why the speed of light is finite,
and finally why String Theory doesn't show us the answer we want,
appear as well distinguished, and apparently unrelated, problems.  
However, in this note I am going to shortly discuss why I do believe
that not only they are deeply related to each other, but that
they may find a common solution into a theoretical framework that
goes beyond Quantum Mechanics and Relativity, that are there lifted to
a description of physical phenomena which is basically neither of them, 
i.e. neither quantum mechanical, nor special/general relativistic 
(and therefore also not field theoretical), although in appropriate limits
it reproduces the aspects these theories describe.

\vspace{1cm}
\centerline{\it The set up}
\vspace{.5cm}

The question we start with can be formulated as follows: is it possible
that the physical world, as we see it, doesn't proceed from a
``selection'' principle, whatever
this can be, but it is just the result, the
``superposition'', the ``mean value'' of all the
possible configurations, in the most general case and with the most
general meaning? May the history of the Universe be viewed somehow as a
path through these configurations, and what we call time ordering an
ordering through the inclusion of sets, so that the configuration at a
certain time is characterized by its containing as subsets all previous
configurations, whereas configurations which are not 
contained belong to the future of the Universe? What is the
meaning of ``configuration'', and
how are then characterized configurations, in order to say which one is
contained and which not?

Following \cite{assiom}, 
let us call configuration a particular way of distributing the most
elementary attributes we can think about,
``cells'' into a space of cells.
We may think of it in terms of assigning which cells of a (target)
space are ``occupied'' and which
ones are ``empty''. This problem can
be viewed in geometric terms, via appropriate discretization of the
coordinates of a vector space through the introduction of a minimal length.
A configuration is then a mapping $\psi$ from a space of 
``cells to be assigned'' to a target space of ``empty cells''. A configuration
$\psi(N)$ is an assignment of $N$ cells:
\be
N ~ \stackrel{\psi (N)}{\longrightarrow} ~ 
M_1 \otimes M_2 \cdots M_i \cdots \otimes M_n \, , ~~~~~~ 
\forall \; \left\{ M_i \, , n \right\} \, , 
~ 1 \leq i \leq n \, . 
\ee
Think at a binary string of information; a binary system 
is the simplest and most elementary way of writing an information code, 
the building block of any more complex system.    
We may then view the target space as a multi-dimensional 
collection of vector spaces of binary strings. 
A configuration is a particular choice of a code in this space.
We may introduce a regularization and 
work at finite volume $V$, defined as the
total number of cells of the target space:  $V = \sum_i M_i$, having however
in mind that we will eventually take the infinite volume limit 
$M_i \to \infty$, $\forall i$. Simply, when working at finite volume,
we must have care that,
if $N$ is the number of occupied cells we are going to
distribute in the target space, at any $V$ we must have $V > N$.

Let's now consider the set of all possible configurations at fixed $N$:
$\left\{ \psi (N)  \right\}$. 
It is clear that for any configuration $\psi (N^{\prime})$, $N^{\prime} < N$,
there is a configuration $\psi (N)$ that
contains $\psi (N^{\prime})$ as a subset (in general, there is more
than one configuration $\psi (N)$ 
that contains $\psi (N^{\prime})$ as a subset).
Therefore, we can say that: 
\be
\left\{ \psi (N)  \right\} \supset \left\{ \psi (N^{\prime})  \right\} \, .
\label{inclusion}
\ee
$N$ plays therefore the role of ``time'', and the ordering (\ref{inclusion})
is a time ordering for our system.
The set $\left\{ \psi (N) \right\}$
is the phase space of the configurations at time $N$.

In order to understand what kind of physics comes out, it is
convenient to map this scenario to an isomorphic description in terms
of geometry, in which $N$ is interpreted as the total energy, $E$, 
of a configuration: $E \stackrel{\rm def}{=} N$, $\psi(E) 
\stackrel{\rm def}{=} \psi(N)$.
A distribution of energy through space determines at any point what
kind of curvature the space possesses. For the time being, by geometry
we intend nothing more than the geometry of the distribution of energy,
that is, of the energy density. As I will
mention later, the geometry of the energy distribution corresponds also
to the geometry of space in the sense of General Relativity. Since any
particle, field, wave packet, and physical phenomenon in general,
consists in a particular geometry of space (i.e. shape of spatial
distribution of energy) that evolves in time, it is clear that, through
the interpretation in terms of geometries of spaces of arbitrary
extension and dimension, this set up is potentially able to describe
all what we observe. Indeed, this framework contains also
configurations that do not correspond to smooth geometries in our
ordinary sense.
This perspective reverses the way of looking at things, and raises the
question on whether all what we see and measure, what we interpret as
the real world, the ``objects that
exist'', indeed everything what exists,
are, in their deepest essence, nothing else than distributions of
degrees of freedom, that we \underline{interpret} in terms of space,
particles, and fields.

The \underline{average} ``geometry'' at
any time/energy $E$ is given by the superposition of all the configurations
belonging to $\left\{ \psi (E)   \right\}$, 
weighted by their volume of occupation in the phase space.
Of course, the phase space contains at any time an infinite number of
elements. This is not true at finite volume. This is why it is
convenient to work at arbitrarily large, but finite, volume.
It is then possible to classify the configurations at any time $E$ and
volume $V$ according to the (relative) number of ways they can be
realized; this in turn allows establishing a hierarchy of weights in
the phase space, or, passing to logarithm, of entropies. An 
important observation is that configurations are identified by their group 
of symmetry and,
since we work at finite volume and discrete groups, there are no two 
inequivalent configurations with the same entropy, in the sense that, 
from a physical point of view, equivalent configurations
are \emph{the same} configuration. In other 
words, we can consider working with equivalence classes of configurations, 
labelled by their symmetry group, rather than with single configurations.      
We can then define a ``generating
function'' for the average geometry and all the
observables of the theory:
\be
{\cal Z}_V (E) ~ = ~ \int_V {\cal D} \psi (E) \, {\rm e}^{\; {1 \over k}
\, S }
\, ,
\label{Zmacro}
\ee
where $k$ is the Boltzmann constant, and ${\cal D} \psi$ simply means
the sum over all configurations, the measure of the integral being
$\exp S/k$.
Since any mean value (of observable) is defined through a
logarithmic derivative, overall volume factors cancel and (\ref{Zmacro})
remains well defined also in the infinite volume limit.

Rather evidently, the integral is dominated by the configurations of
highest entropy.  
Indeed, it happens that, at any time (energy) $E$, the dominant
configuration corresponds to three sphere of radius proportional 
to $E$ \footnote{Needless to say, the passage from combinatorics of
``occupation of cells'' to a
physical/geometrical interpretation requires the introduction of
appropriate units and conversion factors: $c$, $\hbar$, $M_{Pl}$.}. 
The ``universe'' is therefore
predominantly a black hole of radius $\propto E$, and curvature/energy density
$\propto 1/E^2$. This looks much like our universe, in which the
matter/cosmological energy density is indeed $1/T^2$, where $T$ 
is the age/radius up
to the horizon of observation, once units are converted through
appropriate powers of the fundamental constants
\footnote{I adopt here the usual convention of omitting for simplicity
dimensional constants and normalization factors
such as $c$, $\hbar$, $M_{\rm Pl}$, $1/2$ etc.}. In our case, $T = E$. 
Perhaps more astonishingly, it is
possible to evaluate the contribution of \emph{all }the other
configurations at time $E$ to the total energy of the universe. It turns
out that all the remaining configurations contribute for a total amount
$\Delta E \sim 1/E$. $E$, the age of the universe, 
can also be written as $\Delta t$, the interval of time during which
the universe of radius $E$ has been produced. That means, the universe is
mostly a classical black hole, plus a
``smearing'' that quantitatively
corresponds to the Heisenberg Uncertainty, $\Delta E \sim 1 / \Delta T$
\footnote{See previous footnote.}. 
This is due to the fact that
the universe is not just given by one configuration, the dominant one,
but by the superposition of all possible configurations, an infinite
number in which many (an infinite number) don't even correspond to a
three dimensional geometry, or even to a geometry in ordinary sense at
all.
This argument can be refined and applied to any observable one may
define: all what we observe is in fact given by a superposition of
configurations, and whatever value of observable quantity we can
measure is smeared around, is given with a certain fuzziness, that
corresponds to the Heisenberg's inequality.
In this framework, the uncertainty relation arises as a
``global'' way of accounting not
simply for our ignorance about the observables, but for the ill
definiteness of these quantities in themselves, that exist only
``in the average''. For instance,
three-dimensional space itself is such an
``average'' concept, because to the
mean geometry contribute configurations of any dimension. As it arises
in this framework, the Heisenberg Uncertainty accounts for the
contribution of all of them.

It is also possible to show that the speed of expansion of the
--~average, three-dimensional~-- universe, that by convention and choice of
units we can call ``$c$'', is also the
maximal speed of propagation of \underline{coherent}, i.e.
non-dispersive, information. In the
limit in which one passes to the continuum and speaks of space, namely
when one speaks of average three-dimensional world, this can be shown
to correspond to the $v=c$ bound of the speed of light\footnote{Here it is
essential that we are talking of coherent information, as tachyonic
configurations also exist in this scenario, which embeds also Quantum
Mechanics.}. Moreover, the geometry of geodesics in this space corresponds to
the one generated by the energy distribution. This means that this
framework ``embeds'' in itself Special and General Relativity \cite{rel}.

The major novelty of this approach lies in the fact that (\ref{Zmacro}) 
says that the world as we observe it is just the superposition of  all
possible configurations. It states somehow a principle of
``intrinsic necessity'' for the
physical world, that does not come out from a selection principle,
whatever this can be, and whatever can be the form we use to express
it, but simply is the whole of what can be. For astonishing this may
be, it seems to pop out precisely the physics we experience.

\vspace{1cm}
\centerline{\it The dynamics}
\vspace{.5cm}

The dynamics implied by (\ref{Zmacro}) is neither deterministic nor
probabilistic: it is rather a
``determined'' evolution, in which
everything doesn't follow classical causality rules, but the rule of
the highest entropy at any time. On the large scale, this produces an
approximately smooth evolution that we can, up to a certain extent,
parametrize through evolution equations. Since the real world is the
superposition of all configurations at a given time, classical
(= central) values have a spreading out. Even if we could perform the
infinite sum (\ref{Zmacro}) in a finite amount of time (something
clearly not possible), we would anyway not know exactly these values.
The reason is that quantities corresponding to our concept of space
(and therefore also position, wave packet), energy, momentum, etc...
are only defined as average quantities, around the dominant
configuration. The sum (\ref{Zmacro}) contains however also
configurations that we cannot interpret in the usual terms of geometry,
particles, fields, etc.
Therefore, for practical purposes, it turns out to be convenient to
accept a certain amount of unpredictability, introduce probability
amplitudes and work in terms of the rules of quantum mechanics. These
appear as precisely tuned to embed the Heisenberg Uncertainties, as we
found in our framework, into a viable framework enabling to have some
control of the unknown, by endowing them with a probabilistic
interpretation.
Within this theoretical framework, we can therefore give an
interesting interpretation of the Heisenberg's Uncertainty Principle,
and consequently of the \underline{necessity} of a quantum description of
the world. From this point of view, quantization appears as a useful
way of parametrizing the fact of being the observed reality a
superposition of an infinite number of configurations.
The spreading of values of observables implicit in the Heisenberg
Uncertainty Principle does not simply express a limitation of 
our possibility of knowledge, but corresponds to the
limit under which observable quantities in themselves are defined.
Beyond the Uncertainty Principle's threshold, space and time in
themselves are not defined, as we are going to count also the
contribution of configurations that don't have an interpretation in
terms of (classical) geometry, energy, space: they are just assignments
of degrees of freedom (units of energy, if one wants) to a target space
of ``units of position''.

The resulting scenario provides us with a theoretical frame that
unifies quantum mechanics and relativity in a description that,
basically, is neither of them: these turn out to be only
approximations, valid in a certain limit, of a more comprehensive
formulation. Therefore, what had been introduced as a combinatorial
game seems to be the appropriate structure for the description of the
physical world.
The scenario that comes out from (\ref{Zmacro}) is
\underline{highly predictive}, in that there is basically no free
parameter, except from the only running quantity, the age of the
universe, in terms of which everything is computed.
Out of the dominant configuration, a three-sphere,
the averaged contribution given by the other configurations to (\ref{Zmacro})
is responsible for the introduction of ``inhomogeneities'' in the universe, 
that give rise to the varied spectrum of energy clusters 
that we interpret as matter and fields, and, through the time
evolution, their interactions. All of them fall
within the corrections to the dominant geometry implied by the Heisenberg's 
inequality. For instance, matter clusters constitute
local deviations of the mean energy/curvature of order 
$\Delta E \sim 1 / \Delta t $, where $\Delta t$ is the typical
time extension
(or, appropriately converted through the speed of light, the space extension)
of the cluster.
The details of the spectrum can be derived through
a string theoretical representation of the 
combinatorial-geometric scenario.

\vspace{1cm}
\centerline{\it String Theory}
\vspace{.5cm}

In this framework, String Theory arises as a consistent quantum theory
of gravity and interactions, which constitutes a useful mapping of the
combinatorial problem of ``distributions of energy
along a target space'' into a continuum space,
endowed with a minimal length, the continuum version of the unit cell.
To this purpose, one must think at string theory as defined in an
always compact, although arbitrarily extended, space. 
In this case, T-duality, as an exact symmetry in the case of toroidal 
compactification,
or as an approximate, ``softly broken'' symmetry in more general 
compactifications,
ensures the existence of an effective minimal length.
Owing to quantization, and therefore the embedding of the Heisenberg's
Uncertainties, the space of all possible string configurations
\footnote{By ``String Theory'' we mean here the unique theory which is supposed
to underly all the different string constructions.
In this sense, a string configuration has to be intended as a, generally
non-perturbative, configuration of which the possible 
perturbative constructions made in terms of heterotic, or type II, or type I
string, represent ``slices'', dual aspects of the same object.
The existence of such a unique string theory is still an hypothesis, although
well supported by several evidences, that we assume to be true.}
``covers'' all the cases of the
combinatorial formulation, of which it provides a mapping into a
probabilistic scenario, useful for practical computations. 
This mapping entails somehow a ``rearrangement'' of degrees of freedom, 
because string theory is a quantum theory in which
fluctuations of the geometry are described in terms of a spectrum
of excitations that we interpret as particles and fields, that can propagate. 
Each string configuration corresponds therefore to a full collection of 
``static'' combinatorial-geometric configurations, of which it provides
a representation in terms of interactions of propagating 
particles and fields.

The configuration of highest entropy in the phase space of string 
configurations, i.e., the dominant string configuration,
corresponds to the most singular
one, that is, the one with the highest degree 
of reduction of the initial symmetry,
obtained through compactification on curved spaces.  
Its identification is not an easy task, as the perturbative
consistency of a construction is not enough: the real spectrum and
physical content can only be analyzed with strong use of
non-perturbative string dualities \cite{spi}.
The result is a non-perturbative configuration in which the
string target space stabilizes into one time coordinate and three space
coordinates which expand with time, while all the remaining coordinates
are twisted, and therefore stuck, at the Planck scale. This scenario
describes a non-accelerated expanding Universe of radius $R \propto T $
and energy density $\sim 1 / R^2$. Therefore, a three sphere, the 
dominant geometric configuration.
The spectrum corresponds to the degrees of freedom of all the known elementary
particles, and their interactions. However, everything appears
described in a non-standard way, as compared to the usual way one expects
things to appear in a field-, and string-, theory analysis.

A first big change of perspective is that, 
since we work on a space-time which is always compact, there is
no invariance under time translations, because any 
progress in time is a progress in the history
of the universe, and therefore a flow toward a configuration with
different total energy, etc. There is \emph{in general} no invariance under
space translations either, because, unless special boundary conditions
are imposed, any displacement implies approaching, or going away, from the
boundary of space. 
The basic absence of invariance under space-time translations
implies, by construction,
a different normalization of string amplitudes, and therefore a
different interpretation of the computed mean values: owing to
the absence of a normalization factor $1/{\cal V}$, where ${\cal V}$, 
the four-volume of space,
corresponds to the volume of the group of translations, 
densities are now lifted
to global quantities. For instance, the so found string vacuum is
non-supersymmetric, with supersymmetry broken at the Planck scale.
Nevertheless, it produces the correct value of the cosmological
constant, in that the so computed $\langle \Lambda \rangle \, = 1$ vacuum
energy expectation value does not correspond to an energy density, but
to a quantity that, in order to be transformed into a density, must be
rescaled by a Jacobian corresponding to a two-volume, the square radius
of space-time. The so produced true density is therefore
$\Lambda = 1 / R^2 \sim H^2$, where $H$ is the Hubble
constant.
The cosmological constant is correctly predicted in its present day
value, and it turns out to be not at all a constant, but it evolves
with the inverse square of the age of the Universe.

\vspace{1cm}
\centerline{\it Masses and couplings}
\vspace{.5cm}

Another major discrepancy with respect to the traditional scenarios
is that here masses are not generated through a Higgs mechanism, but are
related to the size of space. This in turn is related to the age of the
expanding universe. Therefore, masses depend on time.
Roughly speaking, from a technical point of view masses of elementary 
particles arise
as ground Kaluza-Klein modes in a shifted space-time. In an infinitely
extended space-time they would therefore all vanish. Here they are
related to some power of the inverse of the age of the Universe.
Differently from the massive modes arising from twists and shifts
acting on the internal string coordinates, they lie therefore under the
Planck scale, whereas all the degrees of freedom projected out (such as
for instance extra bosons) are stuck at (or above) the Planck scale.
Sub-Planckian masses arise therefore as a Casimir effect, the lowest
energy modes of a quantum scenario in a compact space. Like the masses
of the internal modes, their existence is consistent with string
theory, and does not require Higgs mechanisms to make the theory
renormalizable. This on the other hand means that gauge theory must be
considered only as an approximation, an effective description.
Indeed, in this scenario, apart from electromagnetism there are no
gauge symmetries in strict terms, as, apart from the photon, there are
no massless bosons in the string spectrum. All symmetries except the
electromagnetic one appear as either already broken, by mass shifts
that lift the corresponding bosons to a non-zero mass (the case of 
the $SU(2)$ of the weak interactions), or in the strong coupling regime
(the $SU(3)$ colour symmetry), and can be investigated only
indirectly through non-perturbative string dualities and mappings to
particular limits. There is therefore no explicit low energy
$U(1) \times SU(2) \times SU(3)$ phase. 
On the other hand, this doesn't exist in nature either!

Apart from the technical details of the way
matter and fields degrees of freedom are generated in the string 
representation, the very origin of a varied
spectrum of masses, and couplings, is due to the fact
that particles and fields have a ``width'' in the phase space.
We are used to see the problem in the other way around, namely, by inserting
mass values and couplings into scattering amplitudes, and obtain that
scattering/decay amplitudes are larger the larger is the 
initial-to-final mass ratio, or the stronger is the coupling of the 
interaction.
In our perspective, the size of masses and couplings is a consequence
of the size of the corresponding process in the phase space. Namely, a 
decay that occurs more frequently than another one, has a stronger coupling.
The strength of the coupling is related to the frequency the process occurs.
In a similar way, a heavier particle decays into lighter particles, and 
therefore its phase space ``contains'' also the phase space of the lighter
particles.  
In other words, masses and couplings are somehow 
related to the ``geometrical probability''
of particles and their interactions in the phase space.

As any string configuration represents a collection of static, geometric 
configurations, a spectrum of matter states obtained through a process
of increasing symmetry reduction, such as the one of the highest entropy
string configuration
\footnote{Here it is given as the one of highest entropy 
in the phase space. In Ref.~\cite{spi} this is instead referred
to as the string configuration of minimal entropy, 
and the analogous of the sum (\ref{Zmacro}) 
in the space of string configurations
is written with a minus sign in front of the 
entropy. The reason is that, besides the entropy in the phase space
of all the string configurations, that we may call the ``macroscopic'' entropy,
one can introduce a ``microscopic entropy'', defined as the 
statistical entropy of the system
constituted by the degrees of freedom of the spectrum of a 
string configuration. It is then possible to show that, at any finite $E$,
``macroscopic'' and ``microscopic'' entropy are dual
to each other, in the sense that, up to an additive constant, 
$S_{\rm micro} = -S_{\rm macro}$. Expressions of 
Ref.~\cite{spi} refer to this second
quantity.}, can be viewed as produced by the superposition
of configurations with progressively reduced amount of symmetry.
At any step the symmetry is reduced, a differentiation in the matter/field
spectrum is introduced. 
As the volumes occupied in the whole phase space by configurations
are inversely proportional to the volumes of their symmetry group, 
the mass ratios between former and new 
particles are related to the ratios of the volumes of 
the symmetry group of the configurations that give rise to them.
The relations are not so simple to be quoted in this small note, so I refer
the reader to \cite{spi} for details. 
To give anyway the flavour of what happens, let us consider a simple 
(unphysical) example.
Let us suppose there is a configuration {\it A} with a spectrum
containing four particles with a symmetry $SU(4)$. Consider now
the configuration {\it B} in which the $SU(4)$ symmetry has been broken
to $SU(2)$. 
The resulting configuration, superposition of the two, 
has therefore a symmetry $SU(2) \times SU(2)$, that we
can indicate as $SU(2)_{I} \times SU(2)_{II}$, and two types of particles: 
the two of type ``{\it I}'', with mass, say, $m_I$, and the two of 
type ``{\it II}'', with mass $m_{II} \neq m_{I}$. 
The particles of type {\it I} correspond to the $SU(2)_I$ symmetry, 
subgroup of $SU(4)$, and are present
in both {\it A} and {\it B}, whereas the particles of type 
{\it II} are those of the broken symmetry, 
and are present only in {\it A}.
The mass difference between particles {\it I} and {\it II} is due 
to the fact that, if the configuration {\it A}
has weight $W(A)$ and the configuration {\it B} has weight $W(B)$ 
in the phase space, the particles {\it I}, present
in both {\it A} and {\it B}, will occur a number $W(A) + W(B)$ of times in the 
phase space, whereas the particles {\it II}, being present
only in {\it A}, will occur only a number
of times $W(A)$. The first ones 
will be therefore heavier than the second 
ones by a factor $m_I = m_{II} \times [(W(A) + W(B)) / W(A)] $, 
as it has to be expected by the fact that the first ones ``contain''
as a subgroup also the physics of the second ones.

By arguments of this kind, although more complicated than in this
elementary example, namely, 
by following the pattern of symmetry reduction in the highest entropy
string configuration, we can determine the relative weight in the
phase space of all types of particles, as functions of the relative weight
of symmetry groups. The couplings too are related to these weights.
The relative ratios of the weight in the phase space
of the various symmetry groups are not pure numerical 
coefficients: pure coefficients are obtained when working on the tangent 
space, with algebras instead of groups. 
Indeed, owing to the multiplicative structure
of the phase space, weights do not sum, as we wrote in the simple example of 
above, but rather multiply. As a consequence, the ratios of weights depend
on the volume of the target space of the string configuration, 
i.e., on the age of the universe.
Proceeding in this way, it is possible to uniquely fix
all the ratios of masses, and the gauge couplings, 
as different powers of the age of the universe. 

We stress here that, owing to the properties of the interpretation of the
Heisenberg's uncertainty we gave, 
the spectrum of particles and fields can be inferred by
looking just at the dominant string vacuum, i.e. the most singular one,
being the other configurations ``covered'' by imposing quantization of the
string. In other words, the contribution of string configurations with exotic
particles and symmetry groups reflects in the quantum uncertainty
with which mean energy values, decays, and scattering amplitudes, are known. 

String techniques allow then to explicitly calculate
the mass exponent corresponding to the lightest mass excitation, 
which is produced by the minimal
shift along the space-time coordinates, in the dominant string 
configuration. It corresponds to the lowest mass scale.   
This identification makes
possible to \underline{compute} then all masses and couplings
\footnote{The uncertainty under which the age of the universe 
is known reflects in an uncertainty 
also for the values of couplings and masses. On the other hand, one can 
reverse the argument, and use one mass value in order to fix the age of 
the universe, and then derive all other masses,
and couplings. In \cite{spi} we used the experimental value of the neutron 
mass, because this is related in simple way to the mass of the only 
non-perturbative stable state, neutral for all interactions,
therefore the only true mass eigenstate, at finite time.}. 
Indeed, since ratios of volumes of symmetry groups
are related to the strength of couplings, which in turn determine the size of 
masses, and the couplings themselves, 
a fine evaluation of these quantities 
proceeds through an iterative series of steps of improving approximation, 
in which the values obtained at the first order are plugged again in
the expressions, to obtain second order values, which are further plugged 
in the expressions to obtain the third order, and so on.  
The so approximated results can then be compared with 
the experimental values of masses and couplings, in order to 
test the theoretical predictions. 

Behind the rather cumbersome procedure of calculation, the rationale
is anyway that masses and couplings 
too are a manifestation of the physics as it comes
out from (\ref{Zmacro}), for which any observable/observed quantity
is the result of a superposition of everything possible, weighted according
to the intrinsic statistics of the ways it can occur. The theoretical
scenario expressed by
(\ref{Zmacro}) represents therefore a radical change of perspective with
respect to the way physical phenomena are traditionally looked at.

Owing to the absence of freely adjustable parameters,
any departure from the
experimental value, in just one prediction, can rule out the whole
theoretical scenario. It is therefore rather remarkable 
that the spectrum of
elementary particles and fields precisely contains all the known
elementary particles and bosons, and that within this framework it has
been possible to compute the fine structure constant with an accuracy
of $\Delta \alpha / \alpha \, \sim  {\cal O} (10^{-5})$, 
the electron and proton mass within
$ \Delta m / m \, \sim {\cal O} (10^{-4})$, etc. (see Ref.~\cite{spi}). 
Even though, for technical reasons, not every quantity can be computed
with the same degree of accuracy,
all the predictions so far derived are in agreement with the experimental 
observations: all
the experimentally detected properties of the known world of elementary
particles and high-energy physics are in this way correctly reproduced,
as a necessary and uniquely determined result.
On the other hand,
no Higgs fields are predicted to exist, nor
low-energy supersymmetry. According to this scenario, no
``new physics'' in the common sense
(new elementary particles, gauge bosons of unification groups, etc...)
is expected to show up at a certain high-energy, under-Planckian scale.
The degrees of freedom of the spectrum correspond to all the already known 
elementary particles and fields, and no more.

Along with a remarkable agreement with the
experimental observations, obtained without tuning of free parameters,
this scenario shows therefore also
an almost complete disagreement with common widely accepted
theoretical expectations.

\newpage
%\vspace{1.5cm}
%\centerline{}
%\vspace{.5cm}

The dependence of the masses of elementary particles
and couplings on the age of the Universe is of the order of 
$ \sim 1/T^{\alpha}$, for appropriate exponents
$\alpha$, different for each mass and coupling. This gives a small, almost
negligible rate of change at present day, but it becomes relevant on a
cosmic scale, where it produces detectable effects. Indeed,
(\ref{Zmacro}) describes not just the physics at
present time, but provides us with a cosmological scenario, predicting
the evolution of the Universe. The Universe turns out to expand in a
non-accelerated way; nevertheless, owing to the time dependence of
masses and couplings, and the consequent shift in the observed
wavelengths of distant objects, the expansion \underline{appears}
accelerated like in a ``matter dominated
universe''. Furthermore, the scaling of masses and couplings is
such that cosmological ``early
universe'' conditions like the nucleosynthesis bound,
or the Oklo bound, usually considered model-killers, are easily
satisfied and provide no significant constraint. On the other hand, the
particular scaling of masses and couplings correctly predicts certain
wavelength shifts observed in the emission spectra of Quasars.

Interestingly, within this framework it is also possible to address
questions of interest for other domains of natural science, such as the
evolutionary biology. The time dependence of masses and couplings
results in the time dependence also of the atomic/molecular emission
and absorption spectra, in particular for what concerns the bounds
responsible for the genetic mutation. If one makes the hypothesis that
mutagenesis is mostly caused by the absorption of natural radiation, it
turns out that mutations are highly favoured during the peaks of
resonance between the frequencies of the absorbing molecule and the
emitting one. As a consequence, natural mutagenesis is expected to
occur in temporal ``phases'' that in
this theoretical framework can be up to a certain extent predicted.
In particular, under the hypothesis of radiation produced by
hydrogen-like sources, an assumption justified by the fact that the
universe is constituted for 3/4 by hydrogen, it is
possible to correctly predict the duration of the Eras of the evolution
of the primates, or the Big Eras of life (Paleozoic, Mesozoic,
Cenozoic) \cite{paleo}. 
This approach opens therefore new perspectives also for this
branch of science.

Indeed, one of the most novel aspects of this theoretical approach is
that it goes beyond the usual organization of physical phenomena we
have in mind, into what we consider pertaining to Quantum Mechanics,
and what to Relativity, allowing us to embrace various aspects in a
perspective ``from above''. An
example is the one just mentioned, in which a deeply
``quantum-gravitational-cosmological''
framework is invoked in order to give a possible explanation of
problems of  biology, whose physical aspects we would consider of
complete pertinence of more classical domains of physics.

The scenario implied by (\ref{Zmacro})  in some sense
``unifies'' General Relativity and
Quantum Mechanics, in that it underlies both of them.
Within this framework, it is possible to investigate phenomena like
the behaviour of a quantum system under relativistic conditions. 
For instance, to study black holes in a true quantum gravity scenario, 
or the behaviour of electrons in a complex system such
as a superconductor, i.e. physical systems which are perhaps beyond the
border of the domain of a perturbative quantum mechanical approach.
These topics are the matter of current investigation.

\newpage

\providecommand{\href}[2]{#2}\begingroup\raggedright\endgroup

\end{document}